\documentclass[letterpaper, 10 pt, conference]{ieeeconf}  

\IEEEoverridecommandlockouts                              
\let\labelindent\relax	

\usepackage[utf8]{inputenc}
\usepackage[T1]{fontenc}
\usepackage{amsmath}
\usepackage{amsfonts}
\usepackage{amssymb}
\usepackage{balance}
\usepackage{graphicx}
\usepackage[rightcaption]{sidecap}
\usepackage{import}
\usepackage{textcomp}
\usepackage{xcolor}
\usepackage{bbm}
\usepackage{slashed}
\usepackage{subfigure}
\usepackage{lipsum}
\usepackage{enumitem}
\usepackage{tabularx}

\newcommand{\bbN}{\mathbb{N}}

\newcommand{\bbR}{\mathbb{R}}

\newcommand{\bbZ}{\mathbb{Z}}

\newcommand{\calC}{\mathcal{C}}

\newcommand{\calE}{\mathcal{E}}

\newcommand{\calG}{\mathcal{G}}

\newcommand{\calK}{\mathcal{K}}

\newcommand{\calN}{\mathcal{N}}

\newcommand{\calP}{\mathcal{P}}

\newcommand{\calU}{\mathcal{U}}
\newcommand{\calV}{\mathcal{V}}

\newcommand{\calX}{\mathcal{X}}

\usepackage{amsthm}

\theoremstyle{definition}
\newtheorem{assumption}{Assumption}
\newtheorem{theorem}{Theorem}
\newtheorem{lemma}[theorem]{Lemma}
\newtheorem{proposition}[theorem]{Proposition}
 
\theoremstyle{remark}
\newtheorem{remark}{Remark}

\title{\LARGE \bf
	A Consistency Constraint-Based Approach to Coupled State Constraints in Distributed Model Predictive Control
}

\author{Adrian Wiltz, Fei Chen, Dimos V. Dimarogonas%
\thanks{This work was supported by the ERC Consolidator Grant LEAFHOUND, the Swedish Foundation for Strategic Research (SSF) COIN, the Swedish Research Council (VR) and the Knut och Alice Wallenberg Foundation.}
\thanks{The authors are with the Division of Decision and Control Systems, KTH Royal Institute of Technology, SE-100 44 Stockholm, Sweden
	{\tt\small \{wiltz,fchen,dimos\}@kth.se}.}%
}

\begin{document}

\maketitle
\thispagestyle{empty}
\pagestyle{empty}

\setlength{\abovedisplayskip}{0.08cm}
\setlength{\belowdisplayskip}{0.08cm}

\begin{abstract}
	In this paper, we present a distributed model predictive control (DMPC) scheme for dynamically decoupled systems which are subject to state constraints, coupling state constraints and input constraints. In the proposed control scheme, neighbor-to-neighbor communication suffices and all subsystems solve their local optimization problem in parallel. The approach relies on consistency constraints which define a neighborhood around each subsystem's reference trajectory where the state of the respective subsystem is guaranteed to stay in. Reference trajectories and consistency constraints are known to neighboring subsystems. Contrary to other relevant approaches, the reference trajectories are improved iteratively. Besides, the presented approach allows the formulation of convex optimization problems even in the presence of non-convex state constraints. The algorithm's effectiveness is demonstrated with a simulation.
\end{abstract}


\section{INTRODUCTION}

Since the publication of the first distributed model predictive control (DMPC) schemes \cite{Camponogara2002}, their development became a thriving branch in the research on model predictive control (MPC). The motivation behind the development of DMPC is that centralized MPC \cite{mayne2000} becomes computationally intractable for large-scale systems, and in the case of spatially distributed systems a reliable communication with a central control-unit is difficult to realize \cite{Frejo2012,Hermans2012}.

In~\cite{mueller2017}, the methods by which distributed MPC algorithms compute control input trajectories are classified into four groups: iterative methods, sequential methods, methods employing consistency constraints, and approaches based on robustness considerations. In iterative methods, the local controllers exchange the solutions to their local optimization problems several times among each other until they converge. In sequential approaches, local optimization problems of neighboring subsystems are not evaluated in parallel but one after another. In algorithms based on consistency constraints, neighboring subsystems exchange reference trajectories and guarantee to stay in their neighborhood. Other DMPC algorithms consider the neighbors' control decisions as a disturbance. Examples can be found in \cite{mueller2017}. As remarked in \cite{Negenborn2014}, the task of distributing MPC algorithms is too complex in order to solve it with one single approach. Instead, for various types of centralized MPC problems, distributed controllers have been taylored. A broad collection of notable DMPC algorithms can be found in \cite{Negenborn2014a}. Many available DMPC schemes follow in their proofs of recursive feasibility and asymptotic stability an argumentation similar to \cite{chen1998} for continuous-time systems, and \cite{DeNicolao1998} for discrete-time systems. 

Especially the distribution of MPC problems subject to coupled state constraints turned out to be complicated \cite{Keviczky2006}, and most available DMPC schemes that are capable of handling them cannot avoid a sequential scheme \cite{Kuwata2011,Nikou2019,Richards2007,Trodden2010}. However, sequential schemes have the drawback that the computation of the control input of all subsystems becomes very time-consuming for highly connected networks when each subsystem has a large number of neighbors. A notable exception that does not rely on a sequential scheme can be found in \cite{farina2012}, where a consistency constraint approach is used instead. This admits that even in the presence of coupled state constraints all subsystems can solve their local optimization problem in parallel and still retain recursive feasibility. However, \cite{farina2012} does not allow to modify the reference trajectory once it is established which restricts the possibility to optimize the system's performance significantly.

In this paper, we overcome this limitation. Taking inspiration from \cite{farina2012}, we employ consistency constraints in our approach. However, we allow to vary and further improve an already established reference trajectory at each time-step. For guaranteeing recursive feasibility, we check some conditions and correspondingly update the reference trajectories afterwards. All computations are carried out in a distributed fashion. 
The proposed algorithm only requires neighbor-to-neighbor communication. 
Interestingly, the usage of consistency constraints allows us to formulate the local optimization problems as convex problems even in the presence of non-convex state and coupled state constraints.
In contrast to \cite{farina2012}, we formulate the algorithm for dynamically decoupled systems.

The remainder is structured as follows: In Section~\ref{sec:problem formulation}, we present the partitioned system and the control objective. In Section~\ref{sec:main results}, we define the local optimization problems, the proposed DMPC algorithm, and derive guarantees. In Section~\ref{sec:simulation}, the algorithm's effectiveness is demonstrated, and in Section~\ref{sec:conclusion} some conclusions are drawn.

\paragraph*{Notation} A continuous function $ \gamma: \bbR_{+}\rightarrow\bbR_{+} $ is a $ \calK_{\infty} $ function if $ \gamma(0) = 0 $, it is strictly increasing and $ \gamma(t)\rightarrow\infty $ as $ t\rightarrow\infty $. If the domain of a trajectory $ x[\kappa] $ with $ \kappa \!=\! a\!,\!\dots\!,\!b $, $ a,b\in\bbN $, is clear from the context, we also write $ x[\cdot] $. By $ x[\cdot|k] $ we denote a trajectory that is computed at time-step~$ k $. The short-hand $ x_{i\in\calV} $ is equivalent to $ \lbrace x_i \rbrace_{i\in\calV} $ where $ \calV $ is an index set. The Minkowski sum is denoted by $ \oplus $. Finally, $ \mathbf{0}, \mathbf{1} $ denote vectors of all zeros or ones, and for $ x\in\bbR^{n} $ and $ P\in\bbR^{n\times n} $, we define the weighted norm $ ||x||_{P} = x^T P x $.


\section{Problem Formulation}
\label{sec:problem formulation}

\subsection{System Dynamics and Constraints}
\label{subsec:II.A}
\vspace{-0.1cm}
Consider a distributed system consisting of subsystems $ i\in\calV = \lbrace 1,\dots,|\calV| \rbrace $ which are dynamically decoupled and behave according to the discrete-time dynamics
\begin{align}
	\label{eq:subsystem dynamics}
	x_i[k+1] = f_{i}(x_i[k],u_i[k]), \qquad x_i[k_0] = x_{0,i}
\end{align}
where $ x_i\in\bbR^{n_i} $ and $ u_i\in\bbR^{m_i} $. The dynamics of the overall system are denoted by
\begin{align}
	\label{eq:system dynamics}
	x[k+1] = f(x[k],u[k]), \qquad x[k_0] = x_{0}
\end{align}
with stack vectors $ x \!=\! [x_1^{T},\dots,x_{|\calV|}^{T}]^{T} $, $ u \!=\! [u_1^{T},\dots,u_{|\calV|}^{T}]^{T} $, and $ x_{0} \!=\! [x_{0,1}^{T},\dots,x_{0,|\calV|}^{T}]^{T} $. All subsystems $ i\!\in\!\calV $ are subject to state constraints $ x_i\!\in\!\calX_i\subseteq\bbR^{n_i} $ and input constraints 
\begin{align}
	\label{eq:input constraint}
	u_i\in\calU_i\subseteq\bbR^{m_i}.
\end{align}
Moreover, some of the subsystems are coupled to each other by coupled state constraints. For subsystems $ i $ and $ j $, if there exists a coupled state constraint $ x_i[k]\in\calX_{ij}(x_j[k]) $, then we call subsystem $ j $ a \emph{neighbor} of subsystem $ i $ and write $ j\in\calN_i $, where $ \calN_i $ is the set of all neighboring subsystems of $ i $. We define the state constraints via inequalities, namely
\begin{subequations}
	\label{eq:state constraints}
	\begin{align}
		h_i(x_i) &\leq \mathbf{0} \\
		c_{ij}(x_i,x_j) &\leq \mathbf{0}, \qquad j\in\calN_i
	\end{align}
\end{subequations}
where $ h_i $, $ c_{ij} $ are continuous; $ h_i: \bbR^{n_i} \rightarrow \bbR^{r_i} $ defines the state constraint with $ r_i $ as the number of inequalities defining the state constraint of subsystem $ i $; $ c_{ij}: \bbR^{n_i}\!\times\!\bbR^{n_j} \!\rightarrow\! \bbR^{s_i} $ defines the coupled state constraints with $ s_i $ as the number of inequalities defining the coupled state constraints that couple subsystem $ i $ with $ j $. 
The state constraint sets are defined as
\begin{align*}
	\calX_i &:= \lbrace x_i \in \bbR^{n_i} \, | \, h_i(x_i) \leq \mathbf{0} \rbrace\\
	\calX_{ij}(x_j) &:= \lbrace x_i\in\bbR^{n_i} \, | \, c_{ij}(x_i,x_j) \leq \mathbf{0} \rbrace, \qquad j\in\calN_i
\end{align*}
where $ \calX_{ij}: \bbR^{n_j} \rightarrow \calP(\bbR^{n_i}) $ is a set valued function and $ \calP(\bbR^{n_{i}}) $ denotes the power set of $ \bbR^{n_{i}} $. 

A state $ x_i $ of subsystem $ i $ is called \emph{infeasible} if \eqref{eq:state constraints} is not satisfied. Besides, we assume that neighboring subsystems that are coupled by state constraints both respect these constraints. In particular, we assume that for all subsystems $ i\in\calV $ and their neighbors $ j\in\calN_i $ we have
\begin{align*}
	c_{ij}(x_i, x_j) = c_{ji}(x_j, x_i).
\end{align*}
Otherwise, this would result in an asymmetry in the subsystems capabilities which is beyond the scope of this paper.


\subsection{Network Topology}
\vspace{-0.1cm}
The coupled state constraints define a graph structure on the distributed system under consideration. The graph is given as $ \calG = (\calV,\calE) $ where $ \calE $ is the set of edges defined as $ \calE := \lbrace (i,j) \, | \, j\in\calN_i \rbrace $. 
We assume that $ \calE $ defines the communication links among the subsystems $ \calV $ and hence that neighboring subsystems can communicate with each other. 


\subsection{Control Objective}
\label{subsec:control objective}
\vspace{-0.1cm}

Let $ \xi_i = f_i(\xi_i,u_{\xi_{i}}) $ be a steady state of subsystem $ i $ for a constant input $ u_{\xi_{i}} $ and denote by $ \xi = [\xi_1^T,\dots, \xi_{|\calV|}^T]^T $ the stack vector of all steady states. The control objective is to steer subsystems $ i\in\calV $ to target states $ \lbrace \xi_{i} \rbrace_{i\in\calV} $ which satisfy
\begin{align*}
	\xi \in \Xi \!:=\! \lbrace \xi | \xi_{i}\!\in\!\calX_{i}, \, \xi_{i}\!\in\!\calX_{ij}(\xi_{j}), \, u_{\xi_{i}} \!\in\!\calU_i, \;\; \forall i\!\in\!\calV, \, \forall j\!\in\!\calN_i \rbrace.
\end{align*}



\section{Main Results}
\label{sec:main results}


\subsection{Local Optimization Problems}
\vspace{-0.1cm}
In the proposed DMPC scheme, a subsystem $ i $ predicts a state-trajectory $ x_i[\kappa | k] $ for $ \kappa = k,\dots,k+N $ and a corresponding input-trajectory $ u_i[\kappa | k] $ for $ \kappa = k,\dots,k+N-1 $ based on dynamics \eqref{eq:subsystem dynamics} at every time-step $ k $ minimizing a cost function $ J_i(x_i[\cdot | k], u_i[\cdot | k]) $ where $ N $ denotes the prediction horizon. Input trajectories $ u_{i\in\calV}[\cdot|k] $ are determined such that they satisfy input constraints \eqref{eq:input constraint}, and state trajectories $ x_{i\in\calV}[\cdot|k] $ such that they start in $ x_{i\in\calV}[k] $  measured at time $ k $ and it holds $ x_i[k|k] = x_i[k] $, $ i\in\calV $. The satisfaction of state constraints~\eqref{eq:state constraints} is ensured by \emph{consistency constraints}
\begin{align}
	\label{eq:consistency constraint}
	x_i[\kappa | k] \in x_i^{\text{ref}}[\kappa | k] \oplus \calC_i
\end{align}
where $ i\!\in\!\calV $, $ \kappa \!=\! k,\dots,k\!+\!N $ and $ \calC_i\!\subseteq\! \bbR^{n_i} $ is a closed neighborhood of the origin.
Thereby, subsystem $ i $ guarantees to its neighbors $ j\!\in\!\calN_i $ that its predicted state trajectory $ x_i[\kappa | k] $ always stays in a neighborhood $ \calC_i $ of reference states $ x^{\text{ref}}_i[\kappa | k] $ for $ \kappa \!=\! k,\dots,k\!+\!N $. Reference states $ x_i^{\text{ref}}[\kappa|k] $ are determined at every time-step and communicated to all neighbors $ j\!\in\!\calN_i $. Contrary to other DMPC approaches employing consistency constraints \cite{Dunbar2007,farina2014,farina2012}, here the reference trajectories $ x_{i\in\calV}^{\text{ref}}[\cdot|k] $ need not to satisfy dynamics~\eqref{eq:system dynamics}. 

The local optimization problem of subsystem $ i $ at time-step $ k $ is given as
\begin{align}
	\label{eq:opt criterion}
	J_i^{\ast}(x_i[k]) = \min_{u_i[\cdot | k] \text{ for } \kappa=k,\dots,k+N} J_i(x_i[\cdot | k], u_i[\cdot | k]) 
\end{align}
where
\begin{align*}
	J_i(x_i[\cdot | k], u_i[\cdot | k]) &= \sum_{\kappa = k}^{k+N-1} l_i(x_i[\kappa|k],u_i[\kappa|k]) \\&\qquad+ J_i^f(x_i[k+N|k]) \\
	l_i(x_i[\kappa|k],u_i[\kappa|k]) &= ||x_i[\kappa | k] - \xi_i||_{Q_i} + ||u_i[\kappa|k] - u_{\xi_i}||_{R_i}
\end{align*}
with stage-cost-function $ l_i $, terminal cost-function $ J_i^f $, and quadratic positive-definite matrices $ Q_i, R_i $. By $ x_i^{\ast}[\cdot|k] $ and $ u_i^{\ast}[\cdot|k] $, we denote those trajectories that minimize $ J_i $. The local optimality criterion \eqref{eq:opt criterion} is subject to
\begin{subequations}
	\label{eq:opt constraints}
	\begin{align}
		\label{seq:opt constraints 1}
		\begin{split}
			\qquad x_i[k|k] &= x_i[k] \\
			x_i[\kappa+1|k] &= f_i(x_i[\kappa|k],u_i[\kappa|k])
		\end{split}
		 \\
		\label{seq:opt constraints 2}
		x_i[\kappa|k] &\in x^{\text{ref}}_{i}[\kappa|k]\oplus \calC_i \\
		\label{seq:opt constraints 3}
		u_i[\kappa|k] &\in \calU_i \\
		\label{seq:opt constraints 4}
		x_i[k+N|k]&\in\calX_i^{f}
	\end{align}
\end{subequations}
for $ \kappa=k,\dots,k+N-1 $ and with terminal set $ \calX^{f}_i $. The optimal control input applied by subsystem $ i $ at time $ k $ given current state $ x_i[k] $ is $ \mu_i(x_i[k]) := u^{\ast}_i[k|k] $.

In order to ensure recursive feasibility and that the consistency constraint do not admit states that result in a violation of state constraints \eqref{eq:state constraints}, the reference states $ x^{\text{ref}}_i[\kappa | k] $, $ \kappa = k,\dots,k+N $, are updated in every time-step $ k $. We impose the following assumption on reference states and consistency constraint sets.
\vspace{-0.12cm}
\begin{assumption}
	\label{ass:consistency constraint}
	For $ \kappa = k,\dots,k+N-1 $ and $ k\geq k_0 $, reference states $ x^{\text{ref}}_i[\kappa | k] $ are chosen 
	\begin{enumerate}[leftmargin=*, start=1,label={(A1.\arabic*)}]
		\item
		\label{ass:consistency constraint a1}
		such that
		\begin{align}
			\label{eq:ref trajectory state constraint}
			h_i(x_i) \leq \mathbf{0} \quad \forall x_i \in x^{\text{ref}}_i[\kappa|k]\oplus\calC_i
		\end{align}
		and for all $ j\in\calN_i $
		\begin{align}
			\label{eq:ref trajectory coupled state constraint}
			\begin{split}
				c_{ij}(x_i,x_j) \leq \mathbf{0} \quad &\forall x_i \in x^{\text{ref}}_i[\kappa|k]\oplus\calC_i, \\
				&\forall x_j \in x^{\text{ref}}_j[\kappa|k]\oplus\calC_j;
			\end{split}
		\end{align}
		
		\item 
		\label{ass:consistency constraint a2}
		and if $ k>k_0 $ additionally such that
		\begin{align}
			\label{eq:ref trajectory recursive feasibillity constraint}
			x^{\ast}_i[\kappa | k-1] \in x^{\text{ref}}_i[\kappa|k ]\oplus\calC_i.
		\end{align}
	\end{enumerate}
\end{assumption}
Whereas \ref{ass:consistency constraint a1} ensures the satisfaction of~\eqref{eq:state constraints}, \ref{ass:consistency constraint a2} leads to recursive feasibility of the local optimization problems. In Section~\ref{subsec:detremination of reference states}, we show how to satisfy Assumption~\ref{ass:consistency constraint}. 

Moreover, we impose the following standard assumptions on terminal sets $ \calX_i^{f} $ and terminal cost $ J_i^{f} $ \cite{farina2012,gruene2011} in a slightly modified way.
\vspace{-0.12cm}
\begin{assumption}
	\label{ass:terminal set}
	For the terminal sets $ \calX_i^{f} $, the terminal costs $ J_i^{f}: \calX_i^{f} \rightarrow \bbR_{\geq 0} $ with $ i\in\calV $, and a state-feedback controller $ k_i^{\text{aux}}(x_i) $, we assume that
	\begin{enumerate}[leftmargin=*, start=1,label={(A2.\arabic*)}]
		\item 
		\label{ass:terminal set a1}
		all $ \calX_i^{f} $, $ i\in\calV $, are feasible, i.e., there exist $ \alpha_i \in \bbR_{>1} $, $ i\in\calV $, such that $ \alpha_i\calX_i^{f} \subseteq \calX_i $ and $ c_{ij}(x_i,x_j) \leq \mathbf{0} \;\;\forall x_i\in\alpha_i\calX_i^{f}, \forall x_j \in\alpha_j\calX_j^{f}$ for all $ j\in\calN_i $;
		
		\item 
		\label{ass:terminal set a2}
		$ k_i^{\text{aux}}: \calX_i^{f} \rightarrow \calU_i $;
		
		\item 
		\label{ass:terminal set a3}
		$ f_i(x_i,k_i^{\text{aux}}(x_i)) \in\calX_i^{f} $;
		
		\item 
		\label{ass:terminal set a4}
		$ J_i^{f}(f(x_i,k_{i}^{\text{aux}}(x_i))) + l_i(x_i,k_i^{\text{aux}}(x_i)) \leq J_i^{f}(x_i) $.
	\end{enumerate}
\end{assumption}
\vspace{-0.12cm}

By~\ref{ass:terminal set a1}, it is ensured that state constraints \eqref{eq:state constraints} are satisfied for any states in the terminal regions $ \calX_{i\in\calV}^f $ and a neighborhood around them.
We suggest to determine $ J_i^{f} $ and $ \calX_i^{f} $ as outlined in \cite[Remark~5.15]{gruene2011}. 


\subsection{Distributed MPC Algorithm}
\vspace{-0.1cm}
First, initial state trajectories $ x^{\text{init}}_i[\kappa|k_0] $, $ \kappa=k_0,\dots,k_0+N $, and input trajectories $ u^{\text{init}}_i[\kappa | k_0] $, $ \kappa=k_0,\dots,k_0+N-1 $, must be determined for all $ i\in\calV $ such that 
\begin{subequations}
	\label{eq:opt init trajectories}
	\begin{align}
		\label{seq:opt init trajectories 1}
		\begin{split}
			x^{\text{init}}_i[k_0|k_0] &= x_{0,i}\\
			x^{\text{init}}_i[\kappa+1|k_0] &= f_i(x^{\text{init}}_i[\kappa|k_0],u^{\text{init}}_i[\kappa|k_0])
		\end{split}\\
		\label{seq:opt init trajectories 2}
		h_i(x^{\text{init}}_i[\kappa|k_0]) &\leq -\beta_i \mathbf{1} \\
		\label{seq:opt init trajectories 3}
		c_{ij}(x^{\text{init}}_i[\kappa|k_0],x^{\text{init}}_j[\kappa|k_0]) &\leq -\beta_i \mathbf{1}\\
		\label{seq:opt init trajectories 4}
		u^{\text{init}}_i[\kappa|k_0] &\in\calU_i\\
		\label{seq:opt init trajectories 5}
		x^{\text{init}}_i[k_0+N|k_0] &\in \calX_i^{f}
	\end{align}
\end{subequations}
where $ \beta_i \in\bbR_{>0} $. Then, we define initial reference trajectories as $ x_i^{\text{ref}}[\cdot|k_0] := x^{\text{init}}_i[\cdot|k_0] $ for all $ i\in\calV $. Note, that \eqref{seq:opt init trajectories 2} and \eqref{seq:opt init trajectories 3} imply \eqref{eq:state constraints} but are stricter. Similarly, \ref{ass:terminal set a1} ensures that for each state in $ \calX_i^{f} $ there exists a neighborhood whose states satisfy \eqref{eq:state constraints} as well. Hence, there always exist sufficiently small sets $ \calC_i $ for all $ i\in\calV $ such that \eqref{eq:ref trajectory state constraint} and \eqref{eq:ref trajectory coupled state constraint} hold with $ x_i^{\text{ref}}[\kappa|k_0] = x^{\text{init}}_i[\kappa|k_0] $, and such that
\begin{align}
	\label{eq:enlarged terminal constraint set}
	\calX_i^{f}\oplus\calC_i \subseteq \alpha_i\calX_i^{f}.
\end{align}

We call trajectories $ x^{\text{init}}_{i\in\calV}[\cdot|k_0], u^{\text{init}}_{i\in\calV}[\cdot|k_0] $ that satisfy \eqref{eq:opt init trajectories} \emph{initially feasible}. We define $ \calX^{0}_{N} $ as the set of states $ x_{0} $ for which initially feasible trajectories exist when the prediction horizon is $ N $.
	For computing the sets $ \calC_{i} $, $ i\in\calV $, we suggest to start with an initial guess of $ \calC_i $ denoted by $ \hat{\calC}_i $, and check if \eqref{eq:ref trajectory state constraint}, \eqref{eq:ref trajectory coupled state constraint} and \eqref{eq:enlarged terminal constraint set} are satisfied. If these hold, enlarge $ \hat{\calC}_i $ by multiplication with a scalar $ \rho_i>1 $; otherwise, shrink $ \hat{\calC}_i $ by multiplication with $ 1/\rho_i $. Then, we choose $ \calC_i = \rho_i^{\iota_{i}^{\text{max}}} \hat{\calC}_i $ where $ \iota_{i}^{\text{max}}\in\bbZ $ is the largest integer such that \eqref{eq:ref trajectory state constraint}, \eqref{eq:ref trajectory coupled state constraint} and \eqref{eq:enlarged terminal constraint set} are still satisfied. If $ \hat{\calC}_i $ is defined as the intersection of half-spaces (polytopes), and $ h_i $ and $ c_{ij} $ are linear, then \eqref{eq:ref trajectory state constraint}, \eqref{eq:ref trajectory coupled state constraint} and \eqref{eq:enlarged terminal constraint set} can be efficiently evaluated using common libraries for computations with polyhedra, e.g., MPT3 (Matlab) \cite{Herceg2013}, Polyhedra (Julia)~\cite{Legat2021} and Polytope (Python) \cite{Polytope}. In \eqref{eq:ref trajectory state constraint} and \eqref{eq:ref trajectory coupled state constraint}, $ \calC_i $ and $ \calC_j $ can be replaced by outer approximations which often results in more conservative conditions, but allows for an easier evaluation. An example is given in Section~\ref{sec:simulation}. 


The entire distributed MPC algorithm, which comprises initialization steps~1 and the DMPC routine~2, is as follows:
\begin{enumerate}
	\item[1.1.] Set $ k=k_0 $ and determine $ J_i^{f}, \calX_i^{f} $ and $ k_i^{\text{aux}} $ for all $ i\in\calV $.
	\item[1.2.] Compute $ x^{\text{init}}_i[\kappa|k_0]$ for $ \kappa = k_0,\dots,k_0+N $ and $u^{\text{init}}_i[\kappa|k_0] $ for $ \kappa = k_0,\dots,k_0+N-1 $ for all $ i\in\calV $ such that \eqref{eq:opt init trajectories} holds. All subsystems $ i\in\calV $ communicate $ x^{\text{init}}_i[\cdot|k_0], u^{\text{init}}_i[\cdot|k_0] $ to their neighbors $ j\in\calN_i $.
	\item[1.3.] For all $ i\in\calV $, set $ x_i^{\text{ref}}[\kappa|k_0] = x^{\text{init}}_i[\kappa|k_0] $ for $ \kappa=k_0,\dots,k_0+N-1 $ and compute $ \calC_i\subseteq\bbR^{n_i} $ as a closed neighborhood of the origin such that \eqref{eq:ref trajectory state constraint}, \eqref{eq:ref trajectory coupled state constraint} and \eqref{eq:enlarged terminal constraint set} hold.
	All subsystems $ i\in\calV $ communicate $ x_i^{\text{ref}}[\cdot|k_0] $ and $ \calC_i $ to their neighbors $ j\in\calN_i $.
	\item[2.1.] All $ i\in\calV $ measure $ x_i[k] $, solve their local optimization problem \eqref{eq:opt criterion} subject to \eqref{eq:opt constraints} in parallel and thereby determine $ x^{\ast}_i[\kappa|k] $ for $ \kappa=k,\dots,k+N $ and $ u^{\ast}_i[\kappa|k] $ for $ \kappa=k,\dots,k+N-1 $. All $ i\in\calV $ communicate $ x^{\ast}_i[\cdot|k], u^{\ast}_i[\cdot|k] $ to their neighbors $ j\in\calN_i $.
	\item[2.2.] All $ i\in\calV $ apply $ \mu_i(x_i[k]) := u^{\ast}_i[k|k] $.
	\item[2.3.] All $ i\in\calV $ determine reference states $ x_i^{\text{ref}}[\kappa|k+1] $ for $ \kappa=k+1,\dots,k+N $ such that Assumption~\ref{ass:consistency constraint} holds (see Section~\ref{subsec:detremination of reference states}). All $ i\in\calV $ communicate $ x_i^{\text{ref}}[\cdot|k+1] $ to their neighbors $ j\in\calN_i $.
	\item[2.4.] Set $ k \leftarrow k+1$ and go to step~2.1.
\end{enumerate}

\vspace{-0.12cm}
\subsection{Guarantees and Properties of the DMPC Algorithm}
\vspace{-0.1cm}
At first, we observe that the proposed DMPC algorithm ensures the satisfaction of the constraints introduced in Section~\ref{subsec:II.A} which each subsystem has to satisfy.

\vspace{-0.12cm}
\begin{lemma}
	For all $ i\!\in\!\calV $, let $ x^{\ast}_i[\kappa|k] $, $ \kappa\!=\!k,\!\dots\!,k\!+\!N $, and $ u^{\ast}_i[\kappa|k] $, $ \kappa\!=\!k,\!\dots\!,k\!+\!N\!-\!1 $, be the solution to local optimization problem~\eqref{eq:opt criterion} with constraints~\eqref{eq:opt constraints}. Then, constraints~\eqref{eq:input constraint} and \eqref{eq:state constraints} are satisfied by $ x_i^{\ast}[\cdot|k] $ and $ u_i^{\ast}[\cdot|k] $.
\end{lemma}
\vspace{-0.12cm}

\vspace{-0.12cm}
\begin{proof}
	This result is straightforward: \eqref{seq:opt constraints 3} implies input constraint \eqref{eq:input constraint}, consistency constraints \eqref{seq:opt constraints 2} imply state constraints \eqref{eq:state constraints} for $ \kappa = k,\dots,k+N-1 $ due to \ref{ass:consistency constraint a1}, and \eqref{seq:opt constraints 4} and \ref{ass:terminal set a1} together imply \eqref{eq:state constraints} for $ \kappa = k+N $.
\end{proof}
\vspace{-0.12cm}

Next, we prove recursive feasibility of the proposed DMPC algorithm and show that the states $ x_{i\in\calV} $ asymptotically converge to their target states $ \xi_{i\in\calV} $.

\vspace{-0.12cm}
\begin{theorem}
	\label{theorem:main theorem}
	For $ i\in\calV $, let $ x^{\text{init}}_i[\kappa|k_0] $ for $ \kappa=k_0,\dots,k_0+N $ and $ u^{\text{init}}_i[\kappa|k_0] $ for $ \kappa=k_0,\dots,k_0+N-1 $ be initially feasible trajectories that satisfy \eqref{eq:opt init trajectories}. Besides, assume that Assumption~\ref{ass:terminal set} holds. Then, the DMPC algorithm comprising steps~1.1 to~2.4 is recursively feasible, $ \xi $ is an asymptotically stable equilibrium of the closed-loop system
	\begin{align*}
		x[k+1] = f(x[k],\mu_N(x[k])),
	\end{align*}
	on $ \calX_{N}^{0} $  and $ x $ asymptotically converges to $ \xi $.
\end{theorem}
\vspace{-0.2cm}
\begin{proof}
	In a first step, we prove recursive feasibility and thereafter asymptotic stability and asymptotic convergence.
	
	\emph{Recursive Feasibility:} In order to show recursive feasibility, we have to show that for all $ k\geq k_0 $, there exist candidate trajectories $ x^{\text{c}}_i[\kappa|k] $ for $ \kappa=k,\dots,k+N $ and $ u^{\text{c}}_i[\kappa|k] $ for $ \kappa=k,\dots,k+N-1 $ that satisfy \eqref{eq:opt constraints}. Thereby it is ensured, that there always exist feasible solutions to the local optimization problems \eqref{eq:opt criterion}-\eqref{eq:opt constraints}.
	
	First, consider $ k=k_0 $ and choose candidate trajectories
	\begin{subequations}
		\begin{align*}
			x_i^{\text{c}}[\kappa|k_0] &= x^{\text{init}}_i[\kappa|k_0] \quad \text{for } \kappa = k_0,\dots,k_0+N, \\
			u_i^{\text{c}}[\kappa|k_0] &= u^{\text{init}}_i[\kappa|k_0] \quad \text{for } \kappa = k_0,\dots,k_0+N-1
		\end{align*}
	\end{subequations}
	for all $ i\in\calV $ where $ x^{\text{init}}_i[\cdot|k_0], u^{\text{init}}_i[\cdot|k_0] $ denote the initially feasible trajectories determined in step~1.2.
	Since $ x^{\text{init}}_i[\cdot|k_0], u^{\text{init}}_i[\cdot|k_0] $ satisfy \eqref{seq:opt init trajectories 1}, it trivially follows that $ x_i^{\text{c}}[\cdot|k_0], u_i^{\text{c}}[\cdot|k_0] $ satisfy \eqref{seq:opt constraints 1} for all $ i\!\in\!\calV $.
	Because $ x_i^{\text{ref}}[\cdot|k_0] \!\!=\!\! x^{\text{init}}_i[\cdot|k_0] \!\!=\!\! x_i^{\text{c}}[\cdot|k_0] $ and $ \calC_i $ is a closed neighborhood of the origin, it also follows that $ x_i^{\text{c}}[\kappa|k_0] $ satisfies \eqref{seq:opt constraints 2}. The satisfaction of \eqref{seq:opt constraints 3}-\eqref{seq:opt constraints 4} trivially follows from \eqref{seq:opt init trajectories 4}-\eqref{seq:opt init trajectories 5}. Thereby, we have shown that there exists at least one feasible solution to the optimization problem \eqref{eq:opt criterion}-\eqref{eq:opt constraints} for $ k\!=\!k_0 $. 
	
	In a next step, we show the existence of feasible solutions for $ k\!>\!k_0 $. Therefore, consider the candidate trajectories
	\begin{subequations}
		\label{eq:candiate trajectories}
		\begin{align}
		\label{seq:candiate trajectories states}
		x_i^{\text{c}}[\kappa|k] &= 
		\begin{cases}
		x^{\ast}_i[\kappa|k-1] & \hspace{-3.3cm}\text{for } \kappa = k,\dots,k+N-1 \\
		f_i(x^{\ast}_i[\kappa-1|k-1],k_i^{\text{aux}}(x^{\ast}_i[\kappa-1|k-1])) & \\&\hspace{-3.3cm}\text{for } \kappa = k+N
		\end{cases}\\
		\label{seq:candiate trajectories inputs}
		u_i^{\text{c}}[\kappa|k] &= 
		\begin{cases}
		u^{\ast}_i[\kappa|k-1] & \hspace{-0cm}\text{for } \kappa = k,\dots,k+N-2 \\
		k_i^{\text{aux}}(x_i[\kappa|k-1]) & \text{for } \kappa = k+N-1
		\end{cases}
		\end{align}
	\end{subequations}
	for all $ i\in\calV $. 
	Since 
	\begin{align*}
		x_i^{\text{c}}[k|k] &= x^{\ast}_i[k|k-1]\\ 
		&= f_i(x^{\ast}_i[k-1|k-1],u^{\ast}_i[k-1|k-1]) \\
		&= f_i(x_i[k-1],u_i[k-1]) = x_i[k]
	\end{align*}
	and since $ x^{\ast}_i[\cdot|k-1], u^{\ast}_i[\cdot|k-1] $ satisfy \eqref{seq:opt constraints 1}, it follows from \eqref{eq:candiate trajectories} that also $ x^{\text{c}}_i[\cdot|k], u^{\text{c}}_i[\cdot|k] $ satisfy \eqref{seq:opt constraints 1}. Because $ x_i^{\text{c}}[\kappa|k] = x_i^{\ast}[\kappa|k-1] $ for $ \kappa = k,\dots,k+N-1 $ and \ref{ass:consistency constraint a2} holds, $ x^{\text{c}}_i[\cdot|k] $ satisfies \eqref{seq:opt constraints 2}. Besides, as $ u_i^{\text{c}}[\kappa|k+1] = u_i^{\ast}[\kappa|k] $, \eqref{seq:opt constraints 3} is trivially satisfied for $ \kappa = k,\dots,k+N-2 $. Furthermore, $ u_i^{\text{c}}[k+N-1|k] = k_i^{\text{aux}}(x_i^{\ast}[k+N-1|k-1])\in\calU_i $ holds due to \ref{ass:terminal set a2} because $ x_i^{\ast}[k+N-1|k-1]\in\calX_i^{f} $, and the satisfaction of \eqref{seq:opt constraints 3} is shown. At last, because $ x_i^{\text{c}}[k\! +\! N|k] = f_i(x^{\ast}_i[k\! +\! N\!-\!1|k\!-\!1],k_i^{\text{aux}}(x^{\ast}_i[k\!+\!N\!-\!1|k\!-\!1])) $	and due to \ref{ass:terminal set a3}, also $ x_i^{\text{c}}[k + N|k] \in\calX_i^{f} $ holds and \eqref{seq:opt constraints 4} is satisfied.
	Thereby, we have shown that there also exists for all $ k>k_0 $ at least one feasible solution to optimization problem \eqref{eq:opt criterion}-\eqref{eq:opt constraints}, and we conclude that the algorithm is recursively feasible.

	\emph{Asymptotic stability and asymptotic convergence (Sketch):} Using standard arguments \cite{gruene2011}, we can derive that $ V_i(x_i) := J_i^{\ast}(x_i) $ is a Lyapunov function for~\eqref{eq:subsystem dynamics} by showing that
	\begin{align*}
		V_i(x_i[k+1])-V_i(x_i[k])&\leq - l_i(x_i[k],\mu_N(x_i[k])) 
		\\&\leq ||x_i[k] - \xi_i||_{Q_i} = -\gamma_{V_i}(x_i[k])
	\end{align*}
	where $ \gamma_{V_i} $ is a $ \calK_{\infty} $ function. Moreover, we can show that $ V(x) := \sum_{i\in\calV} V_i(x) $ is a Lyapunov function for the overall system \eqref{eq:system dynamics} on $ \calX_{N}^{0} $. Then, the asymptotic stability of $ \xi $ on $ \calX_{N}^{0} $ and the asymptotic convergence to $ \xi $ follows, cf. e.g. \cite[Theorem~2.19]{gruene2011}.
\end{proof}


\subsection{Determination of Reference States}
\label{subsec:detremination of reference states}
\vspace{-0.12cm}
For $ k=k_0 $, reference states $ x_{i\in\calV}^{\text{ref}}[\cdot|k] $ are determined in step~1.3 of the DMPC algorithm such that Assumption~\ref{ass:consistency constraint} holds. For $ k>k_0 $, we can determine the reference states $ x_i^{\text{ref}}[\kappa|k] $ for $ \kappa=k,\dots,k+N-1 $ as follows:
\begin{enumerate}
	\item[1.] For $ \kappa=k,\dots,k+N-2 $, check if 
	\begin{align}
	\label{eq:determination ref states state constraints}
		h_i(x_i) \leq \mathbf{0} \qquad \forall x_i \in x_i^{\ast}[\kappa|k-1]\oplus\calC_i
	\end{align}
	and
	\begin{align}
	\label{eq:determination ref states coupled state constraints}
		\begin{split}
			&c_{ij}(x_i,x_j) \leq \mathbf{0} \qquad \forall x_i \in x_i^{\ast}[\kappa|k-1]\oplus\calC_i, \\
			&\;\forall x_j \in (x_j^{\ast}[\kappa|k-1]\oplus\calC_j) \cup (x_j^{\text{ref}}[\kappa|k-1]\oplus\calC_j)
		\end{split}
	\end{align}
	for all $ j\in\calN_i $ hold.
	\item[2.] For $ \kappa=k,\dots,k+N-2 $, if \eqref{eq:determination ref states state constraints} and \eqref{eq:determination ref states coupled state constraints} hold for all $ j\in\calN_i $, then set $ x_i^{\text{ref}}[\kappa|k] = x_i^{\ast}[\kappa|k-1] $, else set $ x_i^{\text{ref}}[\kappa|k]=x_i^{\text{ref}}[\kappa|k-1] $. 
	\item[3.] For $ \kappa=k+N-1 $, set $ x_i^{\text{ref}}[\kappa|k] = x_i^{\ast}[\kappa|k-1] $.
\end{enumerate}

\vspace{-0.15cm}
\begin{proposition}
	\label{prop:determination ref states}
	Let Assumption~\ref{ass:consistency constraint} be satisfied at time-step~$ k_0 $ and let $ k>k_0 $. If $ x_i^{\text{ref}}[\kappa|k] $ is determined according to steps~1 to~3 above for $ \kappa=k,\dots,k+N-1 $ and all $ i\in\calV $, then Assumption~1 holds.
\end{proposition}
\vspace{-0.3cm}
\begin{proof}
	This result follows recursively. To start with, consider $ \kappa=k,\dots,k+N-2 $. At first note that the reference states of all subsystems $ i\in\calV $ are chosen such that $ x_i^{\text{ref}}[\kappa|k]\in\lbrace x_i^{\ast}[\kappa|k-1], x_i^{\text{ref}}[\kappa|k-1] \rbrace $. In step~1, each subsystem $ i\in\calV $ checks if $ x_i^{\text{ref}}[\kappa|k] = x_i^{\ast}[\kappa|k-1] $ is a viable choice such that \ref{ass:consistency constraint a1} is fulfilled for any $ x_j^{\text{ref}}[\kappa|k]\!\in\!\lbrace x_j^{\ast}[\kappa|k\!-\!1], x_j^{\text{ref}}[\kappa|k\!-\!1] \rbrace $. Therefore, \eqref{eq:determination ref states state constraints} and \eqref{eq:determination ref states coupled state constraints} imply \eqref{eq:ref trajectory state constraint} and \eqref{eq:ref trajectory coupled state constraint}, respectively, and \eqref{eq:ref trajectory recursive feasibillity constraint} trivially follows from $ x_i^{\text{ref}}[\kappa|k] = x_i^{\ast}[\kappa|k-1] $. Thus, Assumption~\ref{ass:consistency constraint} holds if \eqref{eq:determination ref states state constraints} and \eqref{eq:determination ref states coupled state constraints} hold.
	
	Next, we show that $ x_i^{\text{ref}}[\kappa|k] = x_i^{\text{ref}}[\kappa|k-1] $ fulfills Assumption~\ref{ass:consistency constraint} independently of the satisfaction of \eqref{eq:determination ref states state constraints} and~\eqref{eq:determination ref states coupled state constraints}. As $ x_i^{\text{ref}}[\kappa|k-1] $ satisfied \eqref{eq:ref trajectory state constraint} at time-step $ k-1 $, it does so at $ k $ since \eqref{eq:ref trajectory state constraint} is decoupled. Next, consider any neighbor $ j\in\calN_i $ of subsystem $ i $. If subsystem $ j $ sets $ x_j^{\text{ref}}[\kappa|k] = x_j^{\ast}[\kappa|k-1] $, then \eqref{eq:determination ref states coupled state constraints} holds according to step~2 and the satisfaction of \eqref{eq:ref trajectory recursive feasibillity constraint} follows. If however subsystem $ j $ sets $ x_j^{\text{ref}}[\kappa|k] = x_j^{\text{ref}}[\kappa|k-1] $, the satisfaction of \eqref{eq:ref trajectory coupled state constraint} follows recursively from the previous time-step. The initial existence of reference states $ x_i^{\text{ref}} $ that satisfy \eqref{eq:ref trajectory coupled state constraint} is ensured by the assumption that $ x_i^{\text{ref}}[\kappa|k_0] $ satisfies Assumption~\ref{ass:consistency constraint}.
	At last, \eqref{eq:ref trajectory recursive feasibillity constraint} is satisfied because the consistency constraint \eqref{seq:opt constraints 2} constraining the optimization problem \eqref{eq:opt criterion}-\eqref{eq:opt constraints} at time-step $ k-1 $ implies $ x^{\ast}_i[\kappa | k-1] \in x^{\text{ref}}_i[\kappa|k-1 ]\oplus\calC_i = x^{\text{ref}}_i[\kappa|k ]\oplus\calC_i $ which is equivalent to~\eqref{eq:ref trajectory recursive feasibillity constraint}. Hence, it is shown that Assumption~\ref{ass:consistency constraint} is also fulfilled, even if \eqref{eq:determination ref states state constraints} or \eqref{eq:determination ref states coupled state constraints} do not hold.	
	At last, consider $ \kappa\!=\!k+N-1 $. Since $ x_i^{\ast}[k\!+\!N\!-\!1|k\!-\!1]\in\calX_i^{f} $ according to \eqref{seq:opt constraints 4}, the choice $ x_i^{\text{ref}}[\kappa|k] = x_i^{\ast}[\kappa|k-1] $ ensures the satisfaction of \ref{ass:consistency constraint a1} due to \eqref{eq:enlarged terminal constraint set} and \ref{ass:terminal set a1}. \ref{ass:consistency constraint a2} is trivially satisfied.
\end{proof}
\vspace{-0.12cm}

\vspace{-0.12cm}
\begin{remark}
	The aforementioned libraries for computations with polyhedra can be also used for the evaluation of conditions \eqref{eq:determination ref states state constraints} and \eqref{eq:determination ref states coupled state constraints}. In particular, note that if sets $ \calC_{i} $, $ i\in\calV $, are polytopes, and $ h_{i} $ and $ c_{ij} $, $ j\in\calN_{i}, i\in\calV, $ are linear, then conditions \eqref{eq:determination ref states state constraints} and \eqref{eq:determination ref states coupled state constraints} reduce to simple algebraic inequalities.
	Besides, observe that $ \calC_i $ and $ \calC_j $ in \eqref{eq:determination ref states state constraints} and \eqref{eq:determination ref states coupled state constraints} can be replaced by an outer approximation which results in more conservative conditions but does not change the result of Proposition~\ref{prop:determination ref states}. This is possible because the choice $ x_i^{\text{ref}}[\kappa|k]=x_i^{\text{ref}}[\kappa|k-1] $ always ensures the fulfillment of Assumption~\ref{ass:consistency constraint} independently of \eqref{eq:determination ref states state constraints} and \eqref{eq:determination ref states coupled state constraints} as it can be seen from the proof of Proposition~\ref{prop:determination ref states}. Outer approximations can help to simplify conditions \eqref{eq:determination ref states state constraints} and \eqref{eq:determination ref states coupled state constraints} especially if $ h_{i}, c_{ij} $ are nonlinear. We give an example in the next section.
\end{remark}


\section{Simulation}
\label{sec:simulation}
\vspace{-0.12cm}

We consider three-wheeled omni-directional robots, which behave according to their kinematic model with states $ x_i \!:=\! [p_{i,x}, p_{i,y},\psi_i]^{T} $ where $ p_{i,x} $ and $ p_{i,y} $ denote the position coordinates and $ \psi_i $ the orientation of robot~$ i $. The position of robot~$ i $ is denoted by $ p \!:=\! [p_{i,x},p_{i,y}]^T $. Its dynamics are 
\begin{align*}
	\dot{x}_i = R(\psi_i) \, (B_i^{T})^{-1} \, r_i \, u_i
\end{align*}
where 
\begin{align*}
	R(\psi_i) \!=\! \bigg[
	\begin{smallmatrix}
	\cos(\psi_i) & -\sin(\psi_i) & 0 \\
	\sin(\psi_i) & \cos(\psi_i) & 0 \\
	0 & 0 & 1
	\end{smallmatrix} \bigg], \;
	B_i \!= \!\bigg[
	\begin{smallmatrix}
	0 & \cos(\pi/6) & -\cos(\pi/6) \\
	-1 & \sin(\pi/6) & \sin(\pi/6) \\
	l_i & l_i & l_i
	\end{smallmatrix} \bigg],
\end{align*}
$ l_i $ is the radius of the robot body, $ r_i $ is the wheel radius, and $ u_i = [u_{i,1},u_{i,2},u_{i,3}]^T $ the angular velocity of the wheels.   

In the sequel, we consider three robots which shall move from an initial formation $ x_0 $ to a target formation $ \xi $, where all robots $ i\in\calV $ are subject to connectivity constraints
\begin{align}
	\label{eq:connectivity constraint}
	||p_i - p_j|| \leq d_{\text{max}}, \qquad j\in\calN_i:=\calV\setminus\lbrace i \rbrace
\end{align}
with $ d_{\text{max}} = 2.6 $, and input constraints $  ||u_i||_{\infty} \leq 15 $ where $ ||\cdot||_{\infty} $ denotes the maximum norm. The performance matrices are chosen as $ Q_1 = \text{diag}(100,100,100) $, $ Q_2 = Q_3 = \text{diag}(1,1,50) $, $ R_1 = \text{diag}(1,1,1) $, $ R_2 = R_3 = \text{diag}(5,5,5) $; the terminal cost functions $ J_{i\in\calV}^{f} $ and terminal sets $ \calX_{i\in\calV}^{f} $ are computed via the algebraic Riccati equation as outlined in \cite{gruene2011}. The consistency constraint set is chosen as a box $ \calC_i = \lbrace \eta \in\bbR^{3} \, | \, ||[\eta_1, \eta_2]||_{\infty} \leq c_i \rbrace $ with $ c_i = 0.125 $ for all $ i\in\calV $; since $ \psi_i $ is unconstrained, the third coordinate $ \eta_3 $ is unconstrained as well. A suitable outer approximation of $ \calC_i $ is $ \bar{\calC}_i := \lbrace \bar{\eta}\in\bbR^{3} \, | \, ||[\bar{\eta}_1, \bar{\eta}_2]||\leq \bar{c}_i \rbrace $ with $ \bar{c}_{i} = \sqrt{2}\, c_i $ where the box constraint on the position coordinates is replaced by a circle that encloses it.
By using the outer approximations $ \bar{\calC}_{i} $, we rewrite \eqref{eq:ref trajectory coupled state constraint} more conservatively as 
\begin{align}
	\label{eq:simplified concistency constraint dist condition}
	c_{ij}(x_i,x_j) = ||p_i - p_j|| - d_{\text{max}} \leq - 2\bar{c}_i,
\end{align}
i.e., the satisfaction of \eqref{eq:simplified concistency constraint dist condition} implies the satisfaction of \eqref{eq:ref trajectory coupled state constraint}. Thereby, it is sufficient to check a simple inequality and set valued operations can be avoided. Although \eqref{eq:ref trajectory coupled state constraint} cannot be simplified in all cases as much as in this example, a problem dependent outer approximation of $ \calC_i $ can often still lead to significant simplifications. If terminal sets $ \calX_{i\in\calV}^{f} $ are chosen sufficiently small such that \eqref{eq:simplified concistency constraint dist condition} is satisfied by all $ x_i\in\calX_{i}^{f} $ for all $ i\in\calV $, then \ref{ass:terminal set a1} is satisfied. Furthermore, if $ \beta_i = 2\bar{c}_{i} $ in \eqref{seq:opt init trajectories 3}, then $ x_{i}^{\text{ref}}[\cdot|k_0] = x_{i}^{\text{init}}[\cdot|k_0] $ fulfills also Assumption~\ref{ass:consistency constraint}. We compute initially feasible trajectories $ x^{\text{init}}_{i\in\calV} $ using a sequential MPC approach. 

In the scenario considered in the simulation, three robots solve a formation control task. Starting in the initial formation $ x_{0,1} = [-1,0,0]^T $, $ x_{0,2} = [-3,1,7\pi/4]^T $, $ x_{0,3} = [-3,-1,\pi/4 ]^T $, the robots move to the target formation $ \xi_1 = [\xi_{11},0,\pi]^T $, $ \xi_2 = [1,-1,\pi/4]^T $, $ \xi_3 = [1,1,7\pi/4]^T $ where $ \xi_{11}\in\lbrace 2.0,2.5,2.75,3.0 \rbrace $. The prediction time is chosen as $ T=12s $ and the prediction horizon as $ N = 36 $. We discretize the time-continuous dynamics using a 4-th order Runge-Kutta method with a time-step $ \Delta t = T/N $. 

In order to evaluate the performance of the proposed algorithm with respect to computation time and actual cost, we compare it with two other common DMPC algorithms: (1) DMPC algorithm of \cite{farina2012} with fixed reference trajectories which cannot be updated once they are established; (2) Sequential DMPC as presented in~\cite[Section~2]{Mueller2012} which is based on the sequential scheme proposed in~\cite{Richards2007}. The MPC controllers have been implemented using Casadi \cite{Andersson2019} and Matlab, the simulations have been performed on an Intel Core i5-10310U with 16GB RAM.

\begin{figure}[tp]
	\centering
	\def\svgwidth{0.7\columnwidth}
	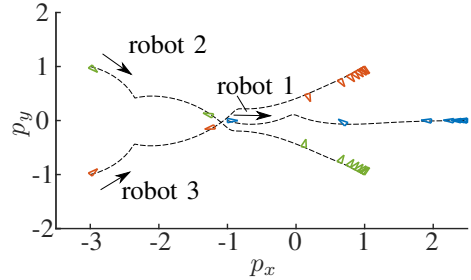
	\caption{Trajectories of robots for $ \xi_{11}=2.5 $. The markers indicate the robots' orientations.}
	\label{fig:trajectories_2p5}
	\vspace{-0.4cm}
\end{figure}

\begin{table}[tp]
	\centering
	\begin{tabularx}{0.5\textwidth}{|>{\centering\arraybackslash\hsize=0.4\hsize\linewidth=\hsize}X||>{\centering\arraybackslash\hsize=1.2\hsize\linewidth=\hsize}X|>{\centering\arraybackslash\hsize=1.2\hsize\linewidth=\hsize}X|>{\centering\arraybackslash\hsize=1.2\hsize\linewidth=\hsize}X|}
		\hline 
		$ \xi_{11} $ & Proposed DMPC & DMPC with fixed reference \cite{farina2012} & Sequential DMPC \cite{Mueller2012}   \\ 
		\hline
		2.0 & 1.00,1.00,1.00 & 0.94,1.42,1.42 & 0.87,0.71,0.71\\
		\hline 	
		2.5 & 1.00,1.00,1.00 & 1.00,1.35,1.32 & 0.93,0.68,0.67 \\
		\hline
		2.75 & 1.00,1.00,1.00 & 1.00,1.18,1.17 & 0.94,0.60,0.59 \\
		\hline
		3.0 & 1.00,1.00,1.00 & 1.01,1.06,1.05 & 0.93,0.53,0.53 \\
		\hline
	\end{tabularx} 
	\caption{Relative actual cost for subsystems 1, 2 and 3.}
	\label{tab:actual cost ratio}
	\vspace{-0.7cm}
\bigskip

	\centering
	\begin{tabularx}{0.5\textwidth}{|>{\centering\arraybackslash\hsize=0.4\hsize\linewidth=\hsize}X||>{\centering\arraybackslash\hsize=1.2\hsize\linewidth=\hsize}X|>{\centering\arraybackslash\hsize=1.2\hsize\linewidth=\hsize}X|>{\centering\arraybackslash\hsize=1.2\hsize\linewidth=\hsize}X|}
		\hline 
		$ \xi_{11} $ & Proposed DMPC & DMPC with fixed reference \cite{farina2012} & Sequential DMPC \cite{Mueller2012}  \\ 
		\hline
		2.0 & 0.0234 & 0.0253 & 0.0935\\
		\hline 	
		2.5 & 0.0237 & 0.0259 & 0.0975 \\
		\hline 	
		2.75 & 0.0233 & 0.0262 & 0.1060 \\
		\hline
		3.0 & 0.0224 & 0.0249 & 0.1405 \\
		\hline
	\end{tabularx} 
	\caption{Average computional times in seconds.}
	\label{tab:computation times}
	\vspace{-1.1cm}
\end{table}

\begin{figure*}[tbp]
	\centering
	\subfigure[$ \xi_{11} = 2.0 $]{
		\def\svgwidth{0.19\textwidth}
		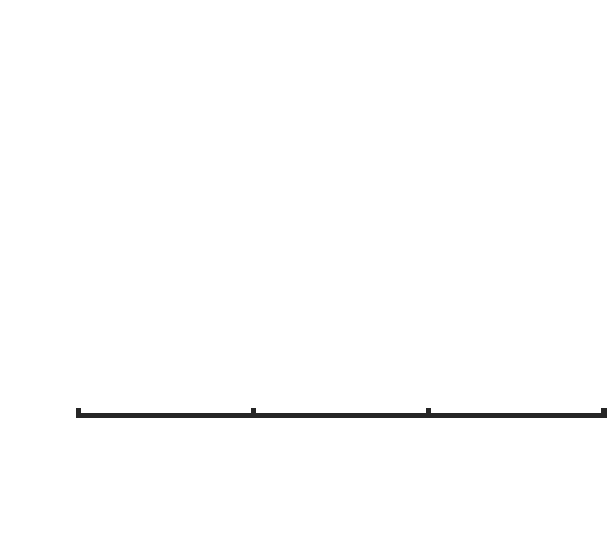
		\label{fig:distances_2p0}
	}
	\subfigure[$ \xi_{11} = 2.5 $]{
		\def\svgwidth{0.19\textwidth}
		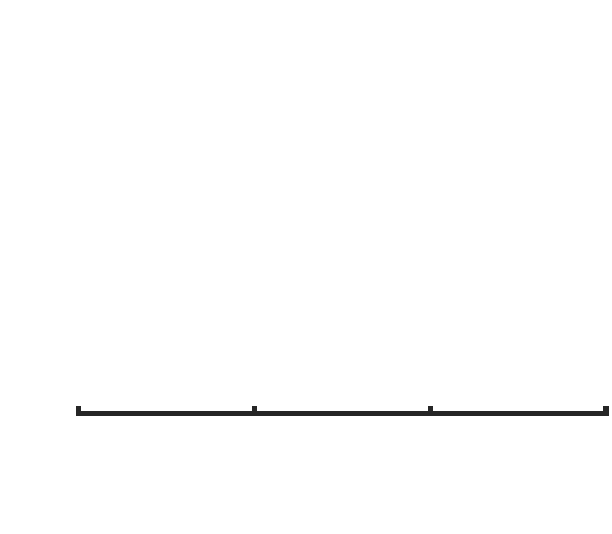
		\label{fig:distances_2p5}
	}
	\subfigure[$ \xi_{11} = 2.75 $]{
		\def\svgwidth{0.19\textwidth}
		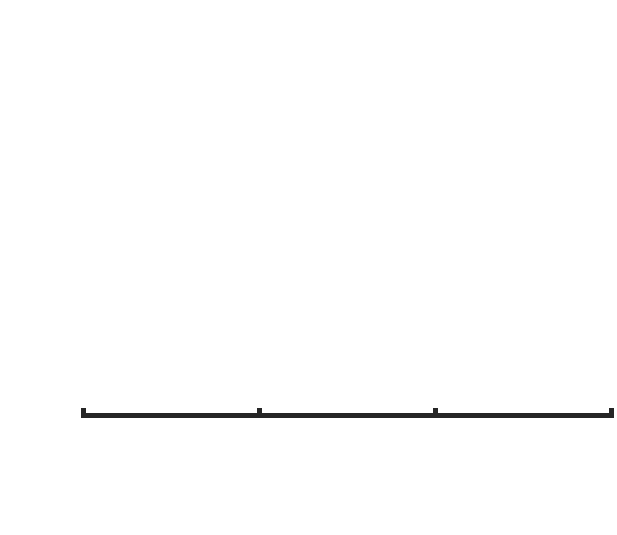
		\label{fig:distances_2p75}
	}
	\subfigure[$ \xi_{11} = 3.0 $]{
		\def\svgwidth{0.19\textwidth}
		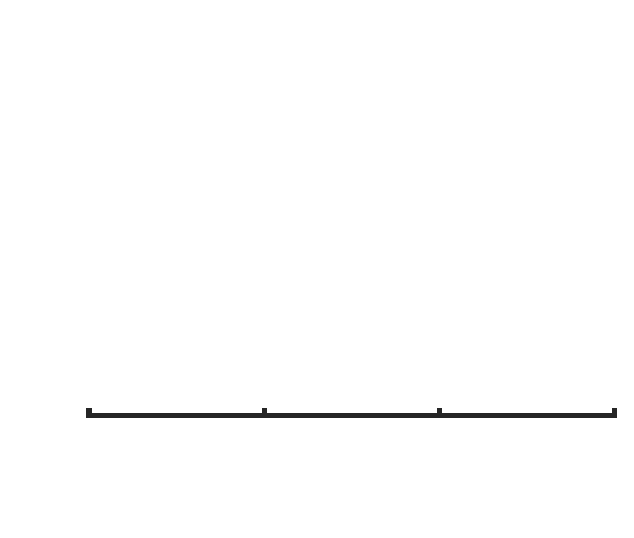
		\label{fig:distances_3p0}
	}
	\caption{Inter robot distances. Distance $ d_{ij} $ denotes the distance between robot $ i $ and $ j $.}
	\label{fig:distances}
	\vspace{-0.5cm}
\end{figure*}

The resulting trajectories of the robots are depicted in Figure~\ref{fig:trajectories_2p5}, the inter-robot distances in Figure~\ref{fig:distances}. From Figure~\ref{fig:distances}, it can be seen that distance constraints \eqref{eq:connectivity constraint} are clearly satisfied. However, simulations using an MPC algorithm that does not take the coupled state constraints \eqref{eq:connectivity constraint} into account resulted for $ \xi_{11}\in\lbrace 2.5, 2.75, 3.0 \rbrace $ in a violation of \eqref{eq:connectivity constraint} which emphasizes the importance of explicitly taking coupled state constraints into account. If the inter-robot distances exceed the dashed-line in Figure~\ref{fig:distances}, then \eqref{eq:determination ref states coupled state constraints} is not satisfied anymore. As a result, the reference state at the respective time is not changed in order to prevent a potential violation of the coupled state constraints (cf. Section~\ref{subsec:detremination of reference states}). This causes some conservativeness which also can be seen from the relative actual costs listed in Table~\ref{tab:actual cost ratio}. For robot $ i $, the actual cost is computed over the simulated time interval as $ J^{a}_{i} = \sum_{\kappa = 0}^{60} ||x_i[\kappa]-\xi_i||_{Q_i} + ||u_i[\kappa]-u_{\xi_{i}}||_{R_{i}} $; the $ i $-th entry in each field of Table~\ref{tab:actual cost ratio} is the actual cost $ J^{a}_{i} $ obtained with the respective DMPC algorithm normed with the actual cost $ J^{a}_{i} $ obtained using the proposed DMPC. Comparing Figure~\ref{fig:distances} and Table~\ref{tab:actual cost ratio} shows, that the more often a subsystem is close to a violation of one of its coupled state constraints and \eqref{eq:determination ref states coupled state constraints} is violated (in this example: the more often the distance curve exceeds the dashed line), the closer is its performance to the performance of fixed reference DMPC (cf. $ \xi_{11}=3.0 $). However, if the distance curves do not exceed the dashed line, or only for small time intervals, the proposed algorithm results in a notably improved performance compared to fixed reference DMPC where the actual costs are 17-42\% higher. 
Most importantly, however, the proposed DMPC computes the control inputs more than 4-6 times faster (Table~\ref{tab:computation times}) than the sequential DMPC. This is due to the parallel evaluation of the local optimization problems in the proposed DMPC and the reduced number of constraints because all state constraints are incorporated into the consistency constraint. This ratio further improves in favor of the proposed DMPC if more subsystems are added because computations of neighboring subsystems must be carried out sequentially.




\section{Conclusion}
\label{sec:conclusion}

We presented a DMPC algorithm that allows for the parallel evaluation of the local optimization problems in the presence of coupled state constraints while it admits to iteratively alter and improve already established reference trajectories at every time-step. For the case of dynamically decoupled systems subject to coupled constraints, we thereby provide a novel DMPC scheme that allows for a faster distributed control input computation compared to sequential DMPC schemes. Guarantees are provided when applying it to the nominal, i.e., undisturbed and known dynamics. In a next step, we plan to extend the presented basic algorithm such that model uncertainties and disturbances can be handled. 

\balance


\bibliographystyle{IEEEtrans}
\bibliography{/Users/wiltz/CloudStation/JabBib/Research/000_MyLibrary}

\end{document}